\documentclass[apj]{emulateapj}

\newcommand{\myion}[2]{\ensuremath{\mathrm{#1}^{#2}}}
\newcommand{\um}{\,\micron}
\newcommand{\neII}{[Ne\,\textsc{ii}]}
\newcommand{\neIII}{[Ne\,\textsc{iii}]}
\newcommand{\neV}{[Ne\,\textsc{v}]}
\newcommand{\sIII}{[S\,\textsc{iii}]}
\newcommand{\siII}{[Si\,\textsc{ii}]}

\newcommand{\oIV}{[O\,\textsc{iv}]}
\newcommand{\feII}{[Fe\,\textsc{ii}]}
\newcommand{\hII}{H\,\textsc{ii}}
\newcommand{\hs}[1]{H$_2$\,S(#1)}

\shortauthors{Smith \textit{et al.}}  
\shorttitle{MIR Spectroscopy of NGC 7331}
\defcitealias{Kennicutt2003}{K03}

\begin{document}

\title{Mid-Infrared IRS Spectroscopy of NGC 7331: A First Look at the
  SINGS Legacy}

\author{%
J.D.T. Smith\altaffilmark{1}, 
D.A. Dale\altaffilmark{2}, 
L. Armus\altaffilmark{3}, 
B.T. Draine\altaffilmark{4}, 
D.J. Hollenbach\altaffilmark{5}, 
H. Roussel\altaffilmark{6},
G. Helou\altaffilmark{6}, 
R.C. Kennicutt, Jr.\altaffilmark{1}, 
A. Li\altaffilmark{1},
G.J. Bendo\altaffilmark{1}, 
D. Calzetti\altaffilmark{7},
C.W. Engelbracht\altaffilmark{1},
K.D. Gordon\altaffilmark{1},
T.H. Jarrett\altaffilmark{6},
L. Kewley\altaffilmark{9}, 
C. Leitherer\altaffilmark{7},
S. Malhotra\altaffilmark{7},
M.J.\ Meyer\altaffilmark{7},
E.J.\ Murphy\altaffilmark{8}
M.W.\ Regan\altaffilmark{7},
G.H.\ Rieke\altaffilmark{1},
M.J.\ Rieke\altaffilmark{1},
M.D.\ Thornley\altaffilmark{10,7}, 
F. Walter\altaffilmark{11}, and
M.G. Wolfire\altaffilmark{12}
}

\altaffiltext{1}{Steward Observatory, University of Arizona, Tucson, AZ  85721}
\altaffiltext{2}{Physics \& Astronomy, Univ. of WY,
Laramie, WY  82071}
\altaffiltext{3}{Spitzer Science Center, Caltech, Pasadena, CA 91125}
\altaffiltext{4}{Princeton University Observatory, Princeton, NJ 08544}
\altaffiltext{5}{NASA Ames Research Center, Moffett Field, CA 94035}
\altaffiltext{6}{Caltech, Pasadena, CA 91101}
\altaffiltext{7}{STScI, Baltimore, MD  21218}
\altaffiltext{8}{Dept. of Astronomy, Yale University, New Haven, CT 06520}
\altaffiltext{9}{Harvard-Smithsonian CfA, Cambridge, MA 02138}
\altaffiltext{10}{Dept. of Physics, Bucknell University, Lewisburg, PA 17837}
\altaffiltext{11}{NRAO, PO Box O, Socorro, NM 87801}
\altaffiltext{12}{Astronomy Dept., Univ. of MD, College Park, MD 20742}
\email{jdsmith@as.arizona.edu}

\begin{abstract}
  The nearby spiral galaxy NGC 7331 was spectrally mapped from 5--38\um\ 
  using all modules of \textit{Spitzer}'s IRS spectrograph.  A strong
  new dust emission feature, presumed due to PAHs, was discovered at
  17.1\um.  The feature's intensity is nearly half that of the
  ubiquitous 11.3\um\ band.  The 7--14\um\ spectral maps revealed
  significant variation in the 7.7 and 11.3\um\ PAH features between the
  stellar ring and nucleus.  Weak \oIV\ 25.9\um\ line emission was found
  to be centrally concentrated in the nucleus, with an observed strength
  over 10\% of the combined neon line flux, indicating an AGN or
  unusually active massive star photo-ionization.  Two \sIII\ lines fix
  the characteristic electron density in the \hII\ regions at
  $n_e$\,$\lesssim$\,$200$\,cm$^{-3}$.  Three detected H$_2$ rotational
  lines, tracing warm molecular gas, together with the observed IR
  continuum, are difficult to match with standard PDR models. Either
  additional PDR heating or shocks are required to simultaneously match
  lines and continuum.
\end{abstract}


\section{Introduction}

The mid-infrared (MIR) spectra of star-forming galaxies are
characterized by fine structure and molecular hydrogen lines, and broad
emission features usually attributed to the large molecular polycyclic
aromatic hydrocarbon (PAH) species, set against a continuum
transitioning from the direct photospheric emission of cool stars at
short wavelengths to the rapidly rising contribution from very small
grains and heated dust toward longer wavelengths.  The Infrared Space
Observatory (ISO) revolutionized the use of both line and band features
in the MIR as diagnostics of star-formation and AGN activity in galaxies
dominated by infrared emission, and found several key relationships
between lines that trace excitation and the hardness of the ultraviolet
radiation field \citep[e.g. \neIII/\neII, ][]{Thornley2000},
star-formation \citep[e.g.  7.7\um\ PAH,][]{Genzel1998} and active
nuclei \citep[e.g.  \neV\ \& \oIV,][]{Sturm2002}.

The \textit{Spitzer Space Telescope} offers for the first time the
sensitivity and spatial resolution necessary to apply the suite of
infrared diagnostics to individual environments in galaxies which span a
full range of interstellar medium (ISM) and IR emission properties.
This is the focus of the \textit{Spitzer} Infrared Nearby Galaxies
Survey \citep[SINGS ---][hereafter
\citetalias{Kennicutt2003}]{Kennicutt2003}, a Legacy program for which
NGC 7331 is the first galaxy observed by \textit{Spitzer}.

NGC 7331 has an IR luminosity typical of normal spiral galaxies with
moderate star-formation activity, with an IR to optical ratio $L_{\rm
  IR}/L_{\rm opt}\sim1$.  By comparison, the extreme starburst M82 has
$L_{\rm IR}/L_{\rm opt}\sim40$, and ultra-luminous infrared galaxies
range in IR activity up to $\sim$400 \citep{Sanders1996}.  Kinematic and
high-resolution studies in both the optical and radio have been unable
to determine whether the galaxy harbors an AGN
\citep[e.g.][]{Filho2002}.

\begin{figure*}[p]
\plotone{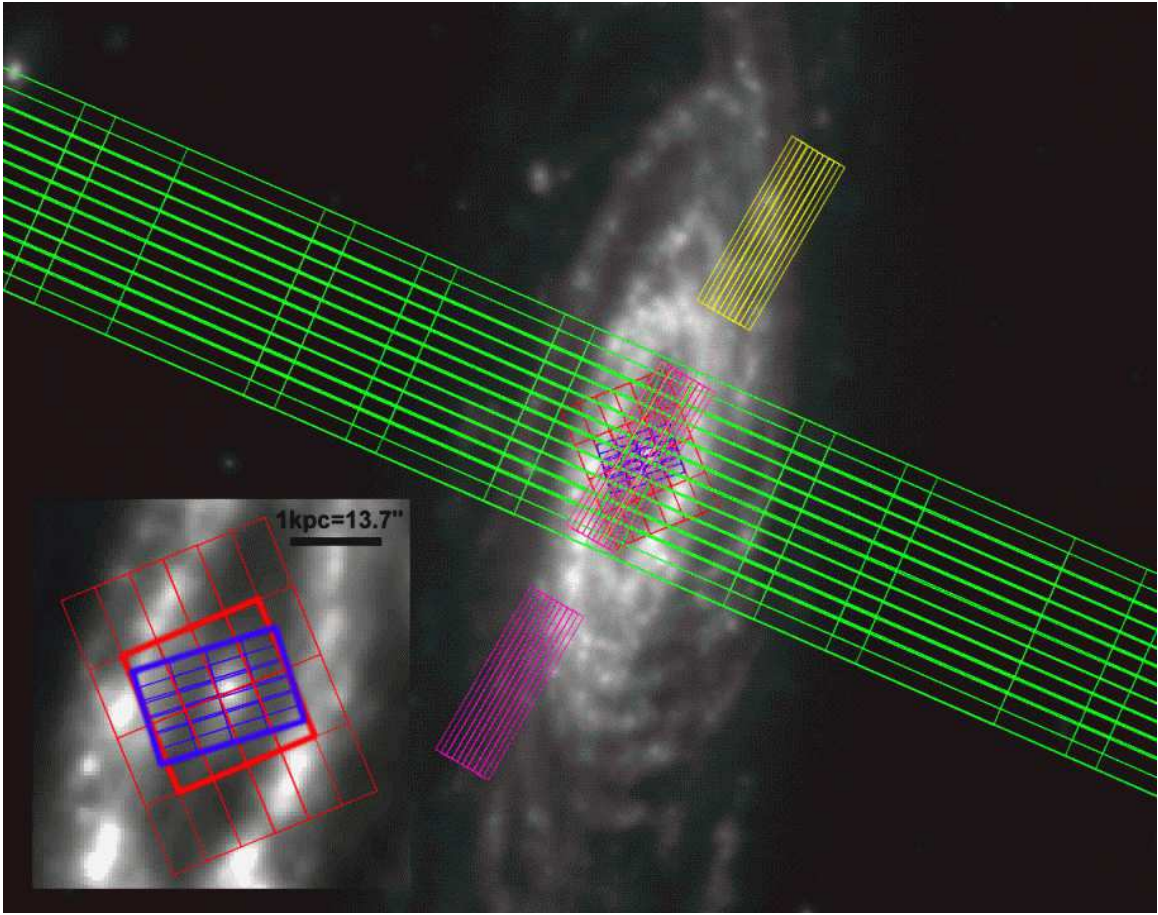}
\caption{An overlay of the IRS footprints on the 8\um\ IRAC image, with
  the nucleus magnified inset, highlighting the corresponding apertures
  used in LH (19--37\um, red) and SH (10--20\um, blue).  The 14--38\um\ 
  LL radial strip is in green (atypically, following the minor axis due
  to unusual scheduling constraints for this galaxy), while the SL
  5--14\um\ regions covered are in magenta and yellow (composed of two
  subslits, with the central area mapped in both subslits).  The
  distance of NGC 7331 is 15.1\,Mpc (1 kpc = 13.66\arcsec).}
  \label{fig:overlay}
\end{figure*}

ISO imaging of NGC 7331 was obtained by \citet{Smith1998}, who found
little variation in the coarse MIR spectral energy distribution traced
in six broad bandpasses from 6--15\um, three of which roughly match
prominent PAH features.  A surprising result from ISO was that the
2.5--11.6\um\ spectral shape is relatively invariant among normal
galaxies, with the appealing explanation that the PAH molecules whose
features dominate this wavelength region are transiently heated by
single photons, and thus produce an emission spectrum roughly
independent of the heating environment \citep{Helou2001}.  The spectral
mapping capabilities of \textit{Spitzer} make it possible to investigate
variations in the PAH emission spectrum \emph{within} individual
galactic environments, and couple these variations using fine structure
and molecular lines to the physical parameters of the ISM, \hII\ and
photo-dissociation regions (PDRs).

\section{Observations and Reduction}

Spectra were obtained using all four modules of the Infrared
Spectrograph \citep[IRS,][]{Houck2004} in \textit{Spectral Mapping
  Mode}.  In this mode, the spacecraft moves in a rectangular raster of
discrete steps, settling at each position before the integrations begin.
For all maps, half-slit step sizes were used.  Since \textit{Spitzer}
cannot roll to alter the position angle of the slits, the mapping
strategy was designed to provide the maximum useful overlap between
modules.  Figure~\ref{fig:overlay} illustrates these areas overlaid on a
SINGS 8\um\ IRAC image \citep[see][in this volume]{Regan2004}, and
indicates the high-resolution apertures used for computing line
parameters.

\begin{figure*}
\plotone{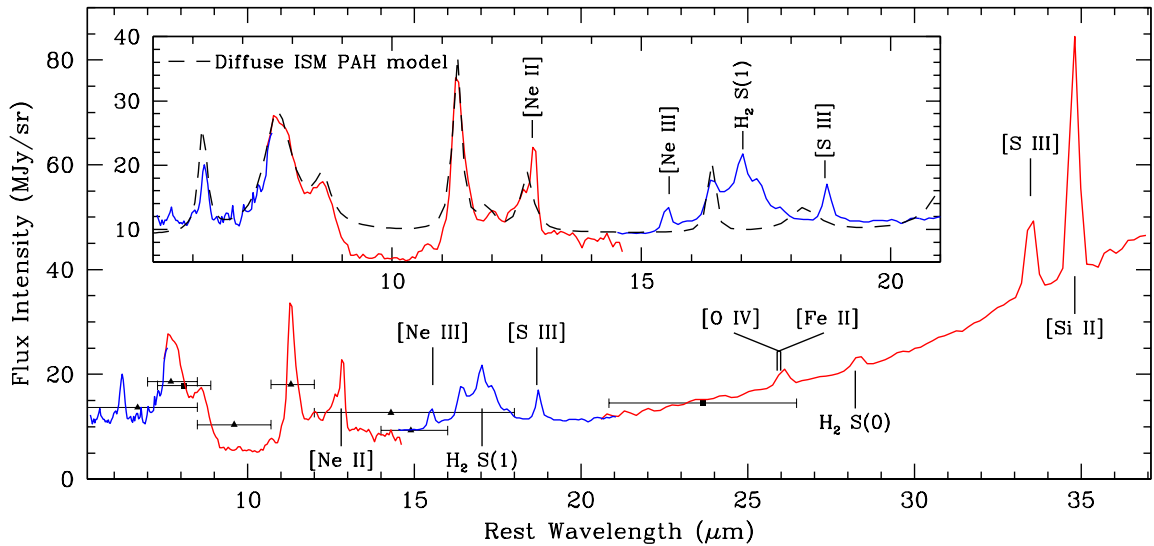}
\caption{
  The low-resolution spectrum of the inner ring and nucleus (central
  magenta rectangle in Fig.~\ref{fig:overlay}).  Alternating colors
  indicate spectra from the four SL and LL orders.  Inset, the expanded
  PAH spectrum with a diffuse ISM PAH + graphite + silicate model
  \citep{Li2001} overlaid as a dashed line, arbitrarily scaled to match
  the 7.7\um\ peak.  The broad 17\um\ complex seen is blended with
  H$_2$~S(1), and contains a sharp feature matching the model at
  16.4\um.  Filled points represent matched photometry with filter
  widths shown, where squares denote IRAC 8\um\ and MIPS 24\um\ 
  \citep{Regan2004}, and triangles denote ISOCAM LW2 (5.0\um), LW6
  (7.0\um), LW7 (8.5\um), LW8 (10.7\um), LW3 (12.0\um), and LW9
  (14.0\um).}
  \label{fig:lowres}
\end{figure*}

Integration times ranged from 14--60s per pointing, with each position
covered 2--4 times, matching the uniform SINGS observational strategy
outlined in \citetalias{Kennicutt2003}.  The two IRS
low-resolution modules, short-low (SL, 5--14\um) and long-low (LL,
14--38\um), provide R=50--100, while the two high-resolution modules,
short-high (SH, 10--20\um) and long-high (LH, 19--38\um), deliver
R$\sim$600.

The individual low-res spectra from each map were assembled into
spectral cubes by \textsc{Cubism}
\citepalias[see][\S6.2]{Kennicutt2003}, which was also used to produce
maps and extract spectra.  The pipeline and calibration products used
were version S9.5.0.  Low-resolution spectra were flux-calibrated using
the standard IRS FLUXCON tables (see Spitzer Observer's
Manual\footnote{\url{http://ssc.spitzer.caltech.edu/documents/som/}},
Chap.\,7), which provide unbiased fluxes only for point sources.  We
derived extended source flux intensities by applying three corrections:
1) an empirically determined \textit{aperture loss correction function}
which accounts for the pipeline's narrowing, point-source specific
extraction aperture, 2) an estimate of the \textit{slit loss correction
  function} which results from the point-spread function over-filling
the IRS slits, and 3) an estimate of the integrated cross-slit beam
profile.  The LL background was subtracted \textit{in situ} using the
extended off-source map wings, and the SL background was fit and removed
using two off-source spectra.  High-resolution spectra were extracted
directly from pre-flatfield data and flux calibrated using stellar
standards via an empirically derived relative spectral response
function.  This resulted in an absolute calibration accuracy of 25\%.

\section{Results}

\subsection{The Spatially Varying PAH Spectrum and a Broad 17\um\ Feature}

\begin{figure}
\centering
\leavevmode
\includegraphics[angle=270,width={.85\linewidth}]{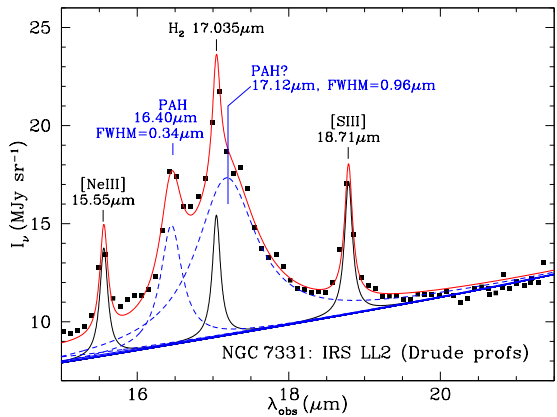}
\caption{
  A fit to the 17\um\ feature, including unresolved \neIII, \sIII, and
  \hs{1} in black, as well as two Drude-profile PAH components in blue:
  the previously discovered 16.4\um\ feature, and a new, much stronger
  17.1\um\ band.
  \label{fig:pah_17um}}
\end{figure}

The entire 5--38\um\ low-resolution spectrum extracted from a
58\arcsec$\times$17\arcsec\ region, covering both the nucleus and a
portion of the ring, is shown in Fig.~\ref{fig:lowres}.  Most of the
fine structure and H$_2$ lines are detected in the low-resolution
spectrum.  The complete PAH spectrum from 5--20\um\ is inset, with an
untailored diffuse ISM PAH + graphite + silicate model of \citet{Li2001}
overlaid.  The model matches the four main PAH bands, at 6.2, 7.7, 8.6,
and 11.3\um, reasonably well.  The feature at 12.7\um\ is contaminated
by \neII\ 12.8\um, but is also in good agreement with the model when the
\neII\ line flux is considered.  Broad 10\um\ silicate absorption is
likely present, as the continuum between 7.7\um\ and 11.3\um\ features
lies below that inferred from bracketing band-free regions, most easily
seen in the strong departure from the scaled model in this region.

Of particular interest is the bright, broad complex at 17\um, blended
with \hs{1}.  The PAH model reproduces the narrow peak on the blue wing
of the feature, but does not predict the large width observed, which is
approximately equal to that of the 7.7\um\ PAH band.  Further peaks
modeled at 18 and 21\um\ are not evident in the spectra.  Deuterated
PAHs are expected to have C-D out-of-plane bending modes which emit in
the 14--16\um\ region, but the observed 17\um\ feature falls longward of
these and is \emph{much} stronger.  It may be due to PAH C-C-C bending
modes \citep[e.g.][]{Allamandola2003}.

\begin{figure*}
\plotone{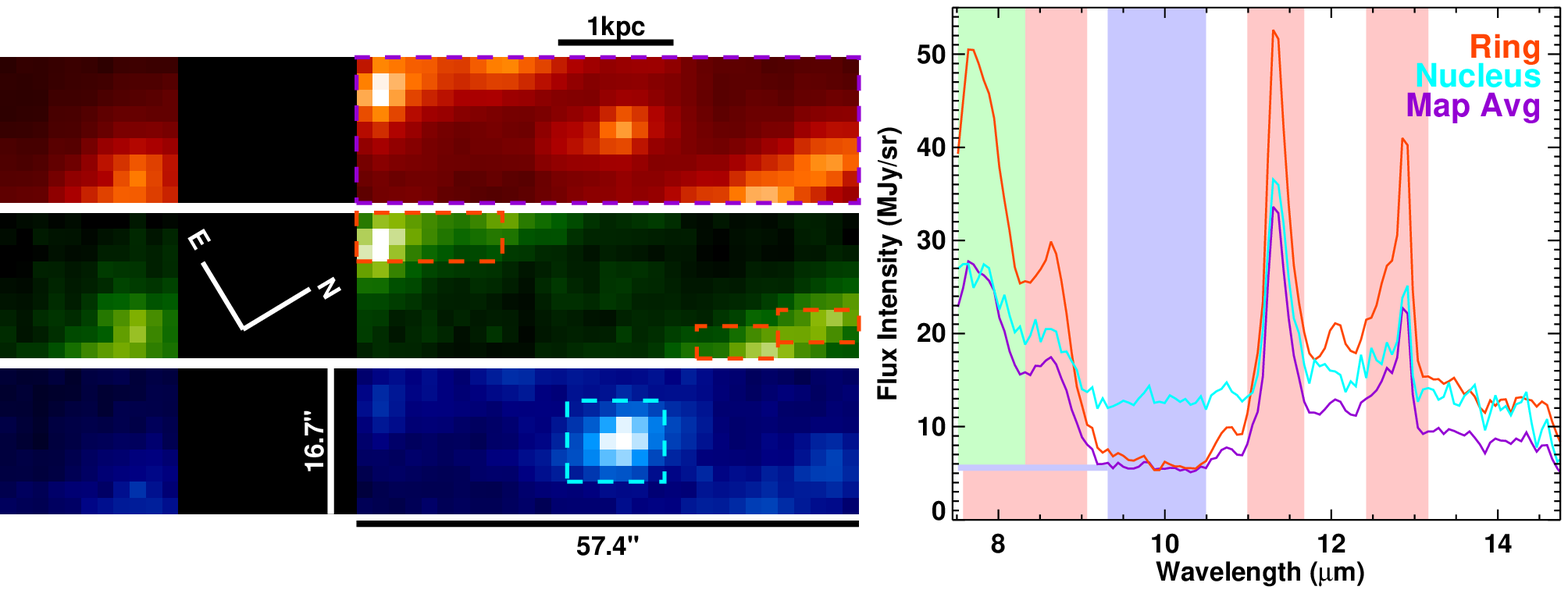}
\caption{
  Three maps created by summing over the SL order 1 (7.5--14.7\um)
  spectral cube, covering the nuclear and a portion of the flanking
  regions illustrated in Fig.~\ref{fig:overlay}, are shown at top.  In
  red (top panel), a sum of the main PAH features at 7.7, 8.6, 11.3 and
  12.7\um; in green (second panel), the continuum-subtracted 7.7\um\ PAH
  feature map; and in blue (third panel) a pure continuum map centered
  at 10\um.  The colored shading in the plot below illustrates the
  spectral regions which were used to create the three maps, including
  10\um\ continuum subtraction in the 7.7\um\ map.  The spectra (bottom
  panel), show three separate extractions in different regions outlined
  with dashed lines in the maps: an average spectrum over the nucleus
  and ring (purple, lowest curve), a ring-only spectrum (orange, highest
  curve), and a nuclear spectrum (cyan, middle curve).
  \label{fig:threemapspec}}
\end{figure*}

\citet{Moutou2000} found a strong, narrow but resolved emission feature
at 16.4\um\ in ISO SWS spectra of the bright reflection nebula NGC 7023,
the Orion bar, and on the surface of a Galactic molecular cloud. Their
spectra showed no other features between 15.2 and 17.2\um.  IRS spectra
of NGC 7023 reveal a series of blended peaks at 16.4, 17.4, and 17.8\um\ 
which vary strongly across the nebula \citep[see][in this
volume]{Werner2004}.  NGC 7331 is the first extragalactic source in
which this feature has been discovered.

In Fig.~\ref{fig:pah_17um}, we fit the blended 17\um\ feature as a
combination of three unresolved lines matching SH identifications, and
two PAH bands --- one as previously reported at 16.4\um, and a new, much
stronger feature at 17.1\um.  Removing the line contaminants, the
combined 16.4\,+\,17.1\um\ feature strength is
$1.5\times10^{-7}$\,W\,m$^{-2}$\,sr$^{-1}$, fully 49\% of the 11.3\um\ 
feature strength, making it a potentially important contributor to the
total MIR luminosity, and a necessary consideration in the
interpretation of photometric redshift and luminosity distributions
based on 24\um\ MIPS or longer imaging.

Using the MIR spectral cubes, we can test the spatial variability of PAH
emission on kiloparsec scales.  Figure~\ref{fig:threemapspec}
illustrates three maps created from the SL spectral cube.  The topmost
map (red) is a combination of the 7.7, 8.6, 11.3, and 12.7\um\ PAH
features (as indicated by shading on the spectra at right), and
highlights the mapped portion of the ring and resolved nuclear core.  In
the continuum-subtracted 7.7\um\ map (green, at middle), the nucleus,
with its weak 7.7\um\ PAH feature, nearly vanishes.  In the pure
continuum map at 10\um\ (blue, at bottom), the strong inter-feature
continuum of the nuclear emission is demonstrated.  The three spectra
extracted from ring, nucleus, and both components together, illustrate
the spatial variability of the relative 7.7\um\ and 11.3\um\ PAH
strengths.  Although the apparent strength of the nuclear 7.7\um\ PAH
feature is diluted by excess continuum, the 7.7\um/11.3\um\ ratio
increases by $\sim$40\% from nucleus to ring.  A similar band ratio
increase, correlated with increasing star formation activity, has also
been observed in M31 \citep{Pagani1999}.

\subsection{H$_2$ and Fine Structure Line Emission}

\begin{figure}
\plotone{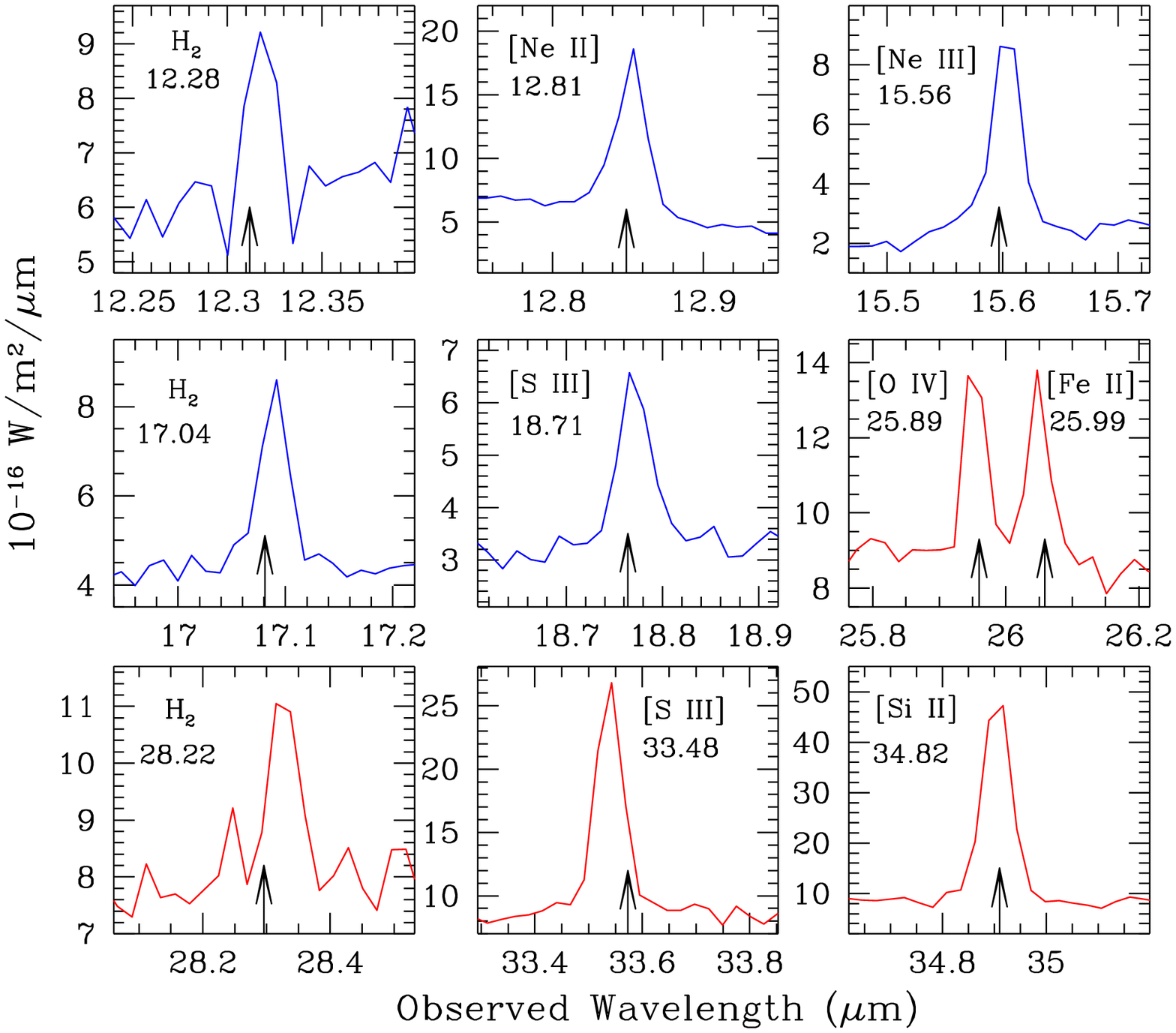}
\caption{Emission lines from the SH and LH spectra extracted from the
  inner nuclear regions (blue and red outlined) indicated in
  Fig.~\ref{fig:overlay}.  Some residual wavelength calibration offsets
  in the pipeline are evident.}
\label{fig:cutouts}
\end{figure}

We detected seven fine structure and three molecular hydrogen lines,
averaged over areas of size 22\arcsec$\times$22.4\arcsec\ (LH) and
23.1\arcsec$\times$15.7\arcsec\ (SH) ($\sim$1--1.5\,kpc on a side at the
15.1\,Mpc distance of NGC 7331, see Fig.~\ref{fig:overlay} inset).  The
fine structure lines detected arise from species ranging in ionization
potential from 8.2\,eV (\myion{Si}{+}) to 55\,eV (\myion{O}{3+}).  The
line parameters are given in Table~\ref{tab:lines}, and individual line
cut-outs are shown in Fig.~\ref{fig:cutouts}.

\begin{deluxetable}{lcr@{\,$\pm$\,}lr@{\,$\pm$\,}lr}
\tablecaption{Observed Line Parameters\label{tab:lines}}
\tablecolumns{7}

\tablehead{
Line &
\colhead{$\lambda_{rest}$} &
\multicolumn{4}{c}{Flux Intensity} &
\colhead{I.P. or $\Delta$E/k\tablenotemark{c}} \\
\colhead{} &
\colhead{} &
\multicolumn{2}{c}{Inner\tablenotemark{a}} &
\multicolumn{2}{c}{Outer\tablenotemark{b}} &
\colhead{}\\
\colhead{} &
\colhead{\um} & 
\multicolumn{4}{c}{$10^{-10}$ W\,m$^{-2}$\,sr$^{-1}$} & 
\colhead{}}
\startdata
\\[-.2in]
\cutinhead{Fine Structure}
\neII   & 12.81  & 237.2& 8.3 & \multicolumn{2}{c}{\nodata} & 21.6\,eV\\
\neIII  & 15.56  & 164.7& 7.5 & \multicolumn{2}{c}{\nodata} & 41.1\,eV\\
\sIII   & 18.68  & 100.7&15.1 & \multicolumn{2}{c}{\nodata} & 23.4\,eV\\
\oIV    & 25.89  &  39.6& 5.5 &   16.2& 4.4                 & 54.9\,eV\\
\feII   & 25.99  &  36.7&11.0 &   29.2& 1.8                 & 7.9\,eV\\
\sIII   & 33.48  & 189.5&19.7 &  251.3&19.7                 & 23.4\,eV\\
\siII   & 34.82  & 463.1&49.1 &  585.9&38.1                 & 8.2\,eV\\   
\cutinhead{Molecular Hydrogen}   
\hs{2}  & 12.28  &  43.8&14.7 & \multicolumn{2}{c}{\nodata} & 1682\,K\\
\hs{1}  & 17.04  & 108.0& 7.5 & \multicolumn{2}{c}{\nodata} & 1015\,K\\
\hs{0}  & 28.22  &  34.2& 5.9 &   43.3& 6.2                 & 510\,K
\enddata
\tablenotetext{a}{Average intensity over $\sim$22\arcsec\ region as
  indicated in Fig.~\ref{fig:overlay}}

\tablenotetext{b}{Weighted average intensity over the entire LH map,
  with 4$\times$ the weight in the regions excluding the nucleus (LH only).}

\tablenotetext{c}{Ionization potential of contributing ion, or
  excitation temperature of upper level state (H$_2$).}
\end{deluxetable}

The 18.71\um\ and 33.48\um\ fine structure lines of \sIII\ are detected.
For $T$\,$\sim$\,$8000$\,K the \sIII\,18.71\um/\sIII\,33.48\um\ line
ratio is insensitive to temperature and provides a density diagnostic.
The density obtained with this diagnostic is
$n_e$\,$\lesssim$\,$200$\,$\mathrm{cm}^{-3}$.  This density corresponds
to a thermal pressure
$p/k$\,$\lesssim$\,$4$\,$\times$\,$10^6$\,cm$^{-3}$K in the \hII\ 
regions.

The pure rotational lines of molecular hydrogen, especially the S(1) and
S(2) lines, which require $T$\,$\gtrsim$\,100\,K for excitation, likely
arise in PDRs surrounding or adjacent to these \hII\ regions, and can be
used to provide joint constraints on the pressure or density and
starlight intensity there.  The observed H$_2$ S(0)/S(1) and S(1)/S(2)
line ratios and intensities can be reproduced by PDR models
\citep[][M.J. Kaufman et al.  2004, in prep.]{Draine2001} with a large
fraction of PDRs in the beam having a similar thermal pressure as the
ionized gas, typically $T$\,$\sim$\,$500$\,K and
$n$\,$\sim$\,$5000$\,cm$^{-3}$ at the surface, with incident FUV fluxes
roughly $G_0$\,$\gtrsim$\,$10^2$ times the local interstellar value.
This is consistent with the strong 20--37\um\ continuum observed in
Fig.~\ref{fig:lowres}, which indicates a component of dust with
temperature 30--60\,K; such warm dust requires
$G_0$\,$\sim$\,$10^2$--$10^3$.

The nucleus and inner edge of the circumnuclear ring in NGC 7331
therefore appear dominated by \hII\ regions and PDRs.  However, our
current PDR models tend to create too much warm dust for the observed
30\um\ continuum.  This may indicate either a lack of understanding of
PDR heating processes \citep[we note similar problems for H$_2$ and CO
rotational emission observed in dense, warm Galactic PDRs,
see][]{Hollenbach1999}, or a contribution of shock heating to the H$_2$
emission.

Also of note is the weak but well-detected \oIV\ line near 26\um\ (see
Fig.~\ref{fig:cutouts}).  \citet{Lutz1998} found that \oIV, with its
high ionization potential of 55\,eV, was nonetheless nearly universally
present as weak emission in galaxies classified as starbursts from the
relative strength of their 7.7\um\ PAH features.  The \myion{O}{3+}
ionization potential is just above the \myion{He}{+} Lyman limit of
54.4\,eV, and as a result \oIV\ is rarely formed in \hII\ regions.  Any
observed \oIV\ therefore requires shocks, the wind-enhanced
far-ultraviolet emission of Wolf--Rayet stars, or an AGN, to provide
sufficient high energy photons \citep[e.g.][]{Schaerer1999}.

When the \oIV/(\neIII+0.44\,\neII) ratio (\myion{O}{3+} to
emissivity-weighted neon emission) is compared to the excitation measure
\neIII/\neII\ (as in \citet{Lutz1998} Fig. 2), NGC 7331 is found to lie
well above the locus of starburst galaxies, with \oIV\ emitting more
than 12\% of the weighted neon line flux (vs.  1\% typical for
traditional starbursts).  That \oIV\ is centrally concentrated is
evident from the relative line intensity in the inner to outer map.
Both the relative strength and tight spatial distribution of \oIV\ 
support the conclusion that a weak AGN provides the ionizing source,
though a central concentration of high-mass stars cannot be ruled out.
The AGN hypothesis is given further credence by noting that preliminary
HST Paschen-$\alpha$ imaging of the center of NGC 7331 indicates little
or no atomic hydrogen line emission from the nucleus (D. Calzetti et al.
2004, in prep.).

\section{Conclusions}
\label{sec:conclusions}

All IRS modules were used to map the inner $\sim$5\,kpc of the nearby
spiral galaxy NGC 7331.  A new band feature at 17.1\um, associated with
the 16.4\um\ PAH feature observed in Galactic nebulae, was found at 49\%
of the 11.3\um\ PAH strength, making it an important contributor to the
total MIR flux.  Among the 7 detected fine structure lines was \oIV,
whose strength and central concentration indicate an AGN or an unusual
concentration of massive stars as the ionizing source.  Two \sIII\ fine
structure lines fix the typical pressure in \hII\ regions to be
$p/k$\,$\lesssim$\,$4$\,$\times$\,$10^6$\,cm$^{-3}$K, and the three
H$_2$ rotational lines require that the bulk of the non-ionizing stellar
radiation be processed through PDRs with similar pressures.  The
traditional PAH features at 7.7 and 11.3\um\ were found to vary in
strength, with the nucleus maintaining strong 11.3\um\ emission relative
to 7.7\um.

\acknowledgments The authors thank J. Houck, V. Charmandaris, and the
IRS team for calibration and validation assistance, and K.  Sellgren for
discussion of NGC 7023's PAH features.  It is also a pleasure to thank
the SINGS liaison scientist, N. Silbermann, and the other members of the
SSC staff for their support during the planning and execution of these
observations.  Support for this work, part of the \textit{Spitzer Space
  Telescope} Legacy Science Program, was provided by NASA through
Contract \#1224769 issued by JPL/Caltech under NASA contract 1407.

\bibliographystyle{apj_special}
\bibliography{n7331}

\end{document}